\documentstyle[11pt,newpasp-sus,twoside]{article} %original

\def\edcomment#1{\iffalse\marginpar{\raggedright\sl#1\/}\else\relax\fi}
\reversemarginpar

\def\kms{km ${\rm s}^{-1}$}

\def\ch2{$\chi^2$}

\def\kms {\hbox{${\rm km\ s}^{-1}$}}

\def\MOLH {\hbox{${\rm H}_2$}}  %H2
\def \HI {H{\sc \,i}}

\def\lapp{\ifmmode\stackrel{<}{_{\sim}}\else$\stackrel{<}{_{\sim}}$\fi}
\def\gapp{\ifmmode\stackrel{>}{_{\sim}}\else$\stackrel{>}{_{\sim}}$\fi}

\begin{document}
\title{Atomic and Molecular Absorption at High Redshift}
\author{S. J. Curran and J. K. Webb}
\affil{School of Physics, University of New South Wales, Sydney NSW 2052, Australia}
\author{M. T. Murphy}
\affil{Institute of Astronomy, Madingley Road, Cambridge CB3 0HA, UK}
\author{Y. M. Pihlstr\"{o}m}
\affil{National Radio Astronomy Observatory, Socorro, NM 87801, USA}
%\author{J. K. Webb}
%\affil{School of Physics, University of New South Wales, Sydney N.S.W. 2052, Australia}   
  
\begin{abstract}
Strong constraints on possible variations in fundamental constants can be
derived from H{\sc \,i} 21-cm and molecular rotational absorption lines
observed towards quasars. With the aim of forming a statistical sample of
constraints we have begun a program of systematic searches for such
absorption systems. Here we describe molecular rotational searches in 25
damped Lyman-$\alpha$ systems where, in many cases, we set optical depth
limits an order of magnitude better than that required to detect the 4 known
redshifted millimeter-wave absorbers. We also discuss the contributory
factors in the detectability of H{\sc \,i} 21-cm absorption, focusing on
possible biases (e.g.~low covering factors) in the currently known sample
of absorbers and non-detections.

\end{abstract}

\section{Introduction}
Quasar absorption lines are powerful probes of possible variations in
fundamental constants over cosmological distances and
timescales. Recent studies of the relative positions of metal-ion
atomic resonance transitions in optical (Keck/HIRES) quasar spectra
are consistent with a smaller fine structure constant in the
intervening absorption clouds over the redshift range $0.2 < z_{\rm
abs} < 3.7$ ({Webb} {et~al.} 1999, Murphy {et~al.} 2003). This
surprising and stubborn result can be cross-checked via several
independent means, either using further optical data from a {\it
different telescope/instrument} or possibly from other quasar
absorption line techniques. Two such techniques are the comparison of
H{\sc i} 21-cm absorption lines with corresponding metal-ion optical
transitions (Cowie \& Songaila 1995) and the comparison of H{\sc i} 21-cm and
molecular rotational (i.e.~millimeter-band) absorption lines
({Drinkwater} {et~al.} 1998, {Carilli} {et~al.} 2000, {Murphy {et~al.} 2001).
However, the paucity of systems
exhibiting H{\sc i} 21-cm and optical/mm-band absorption severely
limits this endeavor. Here we describe our recent attempts to improve
this situation.

\section{Search Strategy and Results}

\subsection{Source selection}

Recently we have commenced a program of scanning the frequency space
towards optically dim millimeter-loud systems in search of possible
absorbers responsible for the visual obscuration (see Murphy,
Curran, \& Webb 2003).  Prior to this, we selected DLAs as targets for
our search since these are the highest column density ($N_{\rm
HI}\gapp10^{20}$ cm$^{-2}$) QSO absorbers known. Also, since there are
many known DLAs with very precisely measured absorption redshifts,
they present the opportunity to form a statistical sample of
absorbers. Additionally, 7 of the 10 known \MOLH-bearing DLAs have high
molecular fractions [$f({\rm H}_2)\sim10^{-2}-10^{-4}$] (see
Reimers et al. 2003), so detection of tracer molecules such as CO and
HCO$^+$ may be possible.

To undertake a systematic search for new high redshift H{\sc i} 21-cm
and molecular absorbers we produced a catalogue of DLAs ({Curran
{et~al.} 2002b)\footnote{A version of this catalogue is continually
updated on-line and is available from
http://www.phys.unsw.edu.au/$\sim$sjc/dla} and shortlisted those which
are illuminated by radio-loud quasars (i.e. those with a measured
radio flux density $>0.1$ Jy). This yielded around 60 DLAs occulting
quasars of sufficient centimetre flux. Of these, 37 have been searched
for 21-cm absorption (see Kanekar \& Chengalur 2003, {Curran {et~al.}
2003). Selecting those of sufficient 12 mm and 3 mm flux gives 18
DLAs which have previously been searched for absorption (see
{Curran {et~al.} 2003). Our recent GBT\footnote{The Green Bank Telescope is
operated by the National Radio Astronomy Observatory.} and
BIMA\footnote{The Berkeley Illinois Maryland Association array is operated
with support from the National Science Foundation.} observations have
increased this number to 30 ({Curran {et~al.} 2004b).

\subsection{Redshifted 21-cm results}

Using the Parkes radio telescope, in January 2002 we searched for
21-cm absorption towards 3 high column density ($N_{\rm
HI}\geq4\times10^{20}$ cm$^{-2}$) DLAs (0432--440, 0438--436 \&
1228--113) strongly illuminated ($S\geq0.35$ Jy) by the background
quasar at $\approx400$ MHz (the frequency of the line at $z\sim2.3$).
Unfortunately, due to severe interference in the 70-cm band, we have
still to fully reduce this data. We note, however, from the published
literature (e.g. Kanekar \& Chengalur 2003) that only half of the 34 DLAs
previously searched have been detected in 21-cm absorption, despite
high neutral hydrogen column densities and strong quasar illumination.
In the literature, the lack of a detection is usually attributed to a
high spin temperature of the neutral atomic gas, based on no other
information than the column density of the Lyman-$\alpha$ line.

After satisfying ourselves that the non-detected DLAs have been
searched sufficiently deeply, we have several reservations about
invoking high spin temperatures, the main being the lack of
information on the covering factor of the quasar (usually
assumed/estimated to be unity). Although little is known of the extent
of the 21-cm absorbing region, through the sizes and morphologies of
the background sources our preliminary results suggest that the quasar
coverage could indeed be a crucial factor in explaining the
non-detections. Any conclusions about a redshift-dependent spin
temperature, and therefore likely DLA host galaxy morphology (Kanekar
\& Chengular 2003), may suffer from a further observational bias:
21-cm absorption detections tend to occur in the $z_{\rm abs}< 1.8$
DLAs which were identified through the Mg{\sc \,ii} 2796/2803 \AA
~doublet, whereas the non-detections at higher $z_{\rm abs}$ are
mostly systems identified through the damped Lyman-$\alpha$ line. If low
covering factor is even partially responsible for the high-$z$
non-detections, the true range in spin temperatures may be far less
than that currently estimated (20 to $>9000{\rm \,K}$), though it may
still be consistent with the hypothesis that $z>2$ DLAs arise in
compact galaxies (see {Kanekar \& Chengalur 2003, {Lanfranchi} \&
{Fria{\c c}a} 2003, Curran et al. 2004a).

\subsection{Redshifted millimeter results}

We have now completed a deep survey of millimeter absorption lines in
DLAs using the Onsala 20-m, Swedish-ESO Sub-millimetre Telescope
({Curran {et~al.} 2002a), the Australia Telescope Compact Array
({Curran {et~al.} 2003) and, most recently, the GBT and BIMA ({Curran
{et~al.} 2004b). From this and the previously published work, 20 DLAs
have been searched to $3\sigma$ limits of $\tau\lapp0.2$ (at 1 \kms
~resolution), i.e. an order of magnitude better than that required to
detect the weakest known mm-absorber ($\tau\approx0.7$ at $\lapp4$
\kms, {Wiklind} \& {Combes} 1994).  The best limits obtained thus far
are $\tau<0.06$ for CO $0\rightarrow1$ ({Takahara} {et~al.}  1987) and
$\tau<0.03$ for HCO$^+$ $0\rightarrow1$ ({Curran {et~al.}
2004b). Using $N_{{\rm H}_{2}}\sim10^4 N_{{\rm CO}}$ ({Wiklind} \&
{Combes} 1998) the former limit gives $N_{{\rm H}_{2}}\lapp1\% N_{{\rm
HI}}$, which is the ratio of the strongest optical H$_2$ detection in
a DLA and so may suggest that in some cases we are close to the CO
detection limit. The HCO$^+$ limit may be a better choice since,
unlike the CO molecule, a constant $N_{{\rm HCO^+}}=2-3\times10^{-9}
N_{{\rm H}_{2}}$ is found over various regimes in the Galaxy (Liszt \&
Lucas 2000). This gives the less impressive limit of $N_{{\rm
H}_{2}}\lapp N_{{\rm HI}}$ for the DLA in question.  It should be
emphasised, however, that converting these limits to molecular
hydrogen column densities is somewhat difficult since Galactic
conversion ratios are based upon dusty, high metallicity systems,
which DLAs are not. We can however use the CO and HCO$^+$ limits to
give column density ratios of $N_{{\rm CO}}\lapp10^{-7} N_{\rm HI}$
and $N_{{\rm HCO^+}}\lapp10^{-9} N_{\rm HI}$ per unit line-width (see
{Curran {et~al.} 2002a). The former value is consistent with the
values of $N_{{\rm CO}}\lapp10^{-8} N_{\rm HI}$ obtained from the
$z>1.8$ redshifted CO electronic transitions ({Black} {et~al.} 1987, {Ge}
{et~al.} 1997, {Lu {et~al.} 1999, {Petitjean} {et~al.} 2002).

\section{Summary}
We have performed deep radio and millimeter integrations of known
high column density absorbers (DLAs) at high redshift in search of
\HI ~21-cm and molecular absorption. From our results in conjunction with those
previously published:
\begin{enumerate}
\item The $50\%$ detection rate of 21-cm absorption may be due to
selection and possible geometrical effects, rather than high gas
temperatures. This would bring the estimated values of the spin
temperatures at high redshift down closer to Galactic values while
still permitting the absorbers to be compact galaxies.
\item We have searched for redshifted millimeter absorption to
sensitivities an order of magnitude better than that required to
detect absorption in the 4 known systems. The CO rotational limits for
DLAs at low redhshift are consistent with those obtained from
electronic transitions redshifted into the optical band at $z>1.8$.
\end{enumerate}

%\newpage

%as usual - paste in .bbl and comment these out

%\bibliographystyle{/home/sjc/styles/apj}
%\bibliography{aa,ref}

\begin{references}
%\expandafter\ifx\csname natexlab\endcsname\relax\def\natexlab#1{#1}\fi

\reference {Black}, J.~H., {Chaffee}, F.~H., \& {Foltz}, C.~B. 1987, ApJ, 317, 442

\reference {Carilli}, C.~L., {Menten}, K.~M., {Stocke}, J.~T., {Perlman}, E., {Vermeulen},
  R., {Briggs}, F., {de Bruyn}, A.~G., {Conway}, J., \& {Moore}, C.~P. 2000,
  PhRvL, 85, 5511

\reference Cowie, L.~L. \& Songaila, A. 1995, ApJ, 453, 596

\reference Curran, S.~J., Murphy, M.~T., Pihlstr\"{o}m, Y.~M., Purcell, C.~R., \& Webb, J.~K. 2004a, in preparation

\reference Curran, S.~J., Murphy, M.~T., Pihlstr\"{o}m, Y.~M., Bolatto, A. D., Bower, G. C,. \& Webb, J.~K.  2004b, in press (astro-ph/0404516)

\reference Curran, S.~J., Murphy, M.~T., Webb, J.~K., \& Pihlstr\"{o}m, Y.~M.
  2003, MNRAS, 340, 139

\reference Curran, S.~J., Murphy, M.~T., Webb, J.~K., Rantakyr\"{o}, F., Johansson, L. E.~B., \& Nikoli\'{c}, S. 2002a, A\&A, 394, 763

\reference Curran, S.~J., Webb, J.~K., Murphy, M.~T., Bandiera, R., Corbelli, E., \& Flambaum, V.~V. 2002b, PASA, 19, 455

\reference {Drinkwater}, M.~J., {Webb}, J.~K., {Barrow}, J.~D., \& {Flambaum}, V.~V. 1998, MNRAS, 295, 457

\reference {Ge}, J., {Bechtold}, J., {Walker}, C., \& {Black}, J.~H. 1997, ApJ, 486, 727

\reference Kanekar, N. \& Chengalur, J.~N. 2003, A\&A, 399, 857

\reference {Lanfranchi}, G.~A. \& {Fria{\c c}a}, A.~C.~S. 2003, MNRAS, 343, 481

%\reference Ledoux, C., Petitjean, P., \& Srianand, R. 2003, MNRAS, astro-ph/0302582

\reference {Liszt}, H.~S. \& {Lucas}, R. 2000, A\&A, 355, 333

\reference Lu, L., Sargent, W. L.~W., \& Barlow, T.~A. 1999, in Highly Redshifted Radio Lines, ed. C.~Carilli, S.~Radford, K.~Menton, \& G.~Langston, Vol. 156 (ASP Conference Series), 132

\reference Murphy, M.~T., Curran, S.~J., \& Webb, J.~K. 2003, MNRAS, 342,
  830

\reference Murphy, M.~T., Webb, J.~K., \& Flambaum, V.~V. 2003, MNRAS, 345, 609

\reference Murphy, M.~T., Webb, J.~K., Flambaum, V.~V., Drinkwater, M.~J., Combes, F., \& Wiklind, T. 2001, MNRAS, 327, 1244

\reference {Petitjean}, P., {Srianand}, R., \& {Ledoux}, C. 2002, MNRAS, 332, 383

\reference {Reimers}, D., {Baade}, R., {Quast}, R., \& {Levshakov}, S. A 2003, A \& A, Accepted (astro-ph/0308432)

\reference {Takahara}, F., {Nakai}, N., {Briggs}, F.~H., {Wolfe}, A.~M., \& {Liszt}, H.~S. 1987, PASJ, 39, 933

\reference {Webb}, J.~K., {Flambaum}, V.~V., {Churchill}, C.~W., {Drinkwater}, M.~J., \& {Barrow}, J.~D. 1999, PhRvL, 82, 884

\reference {Wiklind}, T. \& {Combes}, F. 1994, A\&A, 286, L9

\reference  {Wiklind}, T. \& {Combes}, F. 1998, ApJ, 500, 129

\end{references}
%\expandafter\ifx\csname natexlab\endcsname\relax\def\natexlab#1{#1}\fi
%need for mn2e to work properly

\end{document}